# Overcoming the compression limit of the individual sequence (zero order empirical entropy) using the Set Shaping Theory

Aida Koch, Alix Petit, Christian Schmidt, Adrain Vdberg, Logan Lewis

Abstract: given the importance of the claim, we want to start by exposing the following consideration: this claim comes out more than a year after the article "Practical applications of Set Shaping Theory in Huffman coding" which reports the program that carried out an experiment of data compression in which the coding limit $NH_0(S)$ of a single sequence was questioned. We waited so long because, before making a claim of this type, we wanted to be sure of the consistency of the result. All this time the program has always been public; anyone could download it, modify it and independently obtain the reported results. In this period there have been many information theory experts who have tested the program and agreed to help us, we thank these people for the time dedicated to us and their precious advice. Given a sequence S of random variables i.i.d. with symbols belonging to an alphabet A; the parameter $NH_0(S)$ (the zero-order empirical entropy multiplied by the length of the sequence) is considered the average coding limit of the symbols of the sequence S through a uniquely decipherable and instantaneous code. Our experiment that calls into question this limit is the following: a sequence S is generated in a random and uniform way, the value $NH_0(S)$ is calculated, the sequence S is transformed into a new sequence f(S), longer but with the symbols belonging to the same alphabet, finally we code f(S) using Huffman coding. By generating a statistically significant number of sequences we obtain that the average value of the length of the encoded sequence f(S) is less than the average value of $NH_0(S)$. In this way, a result is obtained which is incompatible with the meaning given to $NH_0(S)$.

## Introduction

The parameter $NH_0(S)$ is considered the average coding limit of the symbols of a sequence S of random variables i.i.d through a uniquely decipherable and instantaneous code. In article [1] we presented a program that performs a data compression experiment which represents a counterexample of this limit.

Regarding this limit, it is important to specify that when we talk about coding an individual sequence (the source is not known) the compressed message must also include the coding scheme which can be attached to the message or as happens in many universal coding, obtainable from the encoded message. Consequently, an inefficiency is created which has various names: the most common is "price of universality" or "inefficiency of entropic coding". Therefore, exceeding this limit involves a reduction in this inefficiency. <u>This clarification serves to make people understand that exceeding this limit does not lead to absurd results such as, for example, being able to compress a random sequence but simply manages to reduce</u>

Contact author: aida.koch445@outlook.com

The value of $NH_0(S)$ is considered the coding limit of the sequence S because $H_0(S)$ is considered as the minimum value of self-information per symbol. Travis Gagie in the article [2] reports a theorem that he uses to support this statement. We will analyze this theorem and its proof and highlight a fundamental criticality that may be the cause of the wrong interpretation of this theorem.

## Zero order empirical entropy and Set Shaping Theory

Let S be a string of length N over the alphabet $A = \{a_i, \ldots \ldots a_h\}$, and let $n_i$ denote the number of occurrences of the symbol $a_i$ inside S. The zeroth order empirical entropy of the string S is defined as:

$$H0(S) = -\sum_{i=1}^{h} n_i/n \log_2 n_i/n$$

The zero-order empirical entropy $H_0(S)$ multiplied by the length of the sequence N is considered to be coding limit of an individual sequence through the fixed code symbols using a uniquely decipherable and instantaneous code [3].

Codes of this type have the characteristic that no codeword is the prefix of another codeword. Therefore, a code is uniquely decodable if it is possible to decode each transmitted character uniquely, without ambiguity. These codes are very important because many theorems such as Shannon's first theorem [4] refer to this type of codes.

The Set Shaping Theory [5][6] has as its objective the study and application in information theory of bijection functions *f* that transform a set $X^N$ of strings of length *N* into a set $Y^{N+K}$ of strings of length *N+K* with *K* and $N \in \mathbb{N}^+, |X^N| = |Y^{N+k}|$ and $Y^{N+K} \subset X^{N+K}$.

The function *f* defines from the set $X^{N+K}$ a subset of size equal to $|X^N|$. This operation is called "Shaping of the source", because what is done is to make null the probability of generating some sequences belonging to the set $X^{N+K}$.

The parameter *K* is called the shaping order of the source and represents the difference in length between the sequences belonging to $X^N$ and the transformed sequences belonging to $Y^{N+k}$[7].

Contact author: aida.koch445@outlook.com

The functions that respect this condition are many but since the goal of this theory is the transmission of data, the function $f_m$ studied is which transforms the set $X^N$ into the set $Y^{N+k}$ composed of the $|X^N|$ strings with less $H_0(x)$ belonging to $X^{N+k}$.

The bijection function $f_m$ is defined as:
$$f_m: X^N \to Y^{N+k}$$
With $K, N \in \mathbb{N}^+$, $|X^N| = |Y^{N+k}|$, $Y^{N+K} \subset X^{N+K}$, $X^{N+K} - Y^{N+K} = C^{N+K}$, $\forall\, y \in Y^{N+K}$ and $\forall\, c \in C^{N+K}$ $H_0(y) < H_0(c)$ and $H_0(y_i) < H_0(y_{i+1})$ $\forall\, y \in Y^{N+K}$.

Given a set $X^N$ which contains all the sequences of length N that can be generated, therefore with dimension $|X^N| = |A|^N$. The Set Shaping Theory tells us that when $|A| > 2$ exists a set $Y^{N+K}$ of dimension $|A|^N$ consisting of sequences having alphabet A and length N+K, in which the average value of $(N+k)H_0(y)$ is less than the average value of $NH_0(x)$ calculated on the sequences belonging to $X^N$.

Since the two sets have the same dimension, it is possible to put the sequences belonging to the two sets into a one-to-one relationship. Consequently, this function would allow us to exceed the limit defined by the parameter $NH_0(S)$ obtaining a result of enormous interest.

## Counterexample

Given a sequence S consisting of i.i.d random variables belonging to an alphabet A of length N, the parameter $H_0(S)$ is considered to be the minimum value of self-information per symbol. Therefore, $NH_0(S)$ is considered to be the coding limit, in which the symbols belonging to the alphabet A are replaced by a uniquely decipherable and instantaneous code.

Consequently, this limit has two fundamental constraints:

The first concerns the type of code with which to encode the symbols of the alphabet A which must be uniquely decipherable and instantaneous.

The second concerns the alphabet which must remain unchanged, in fact $H_0(S)$ is related to the symbols belonging to the alphabet A.

Now, let's build our counterexample starting from these two constraints.

As a coding method we take Hoffman coding [8] which generates uniquely decipherable and instantaneous codes.

Contact author: aida.koch445@outlook.com

We assume as a condition that in our counterexample the alphabet must never change.

Based on these considerations we have developed the following counterexample:

1) Generate a random sequence S with uniform distribution (symbol emission probability 1/|A|) with alphabet A and length N.

2) Calculate the parameter $NH_0(S)$ as defined previously.

3) We perform a transform on sequence S generating a new sequence f(S) with symbols belonging to A (the alphabet is not changed).

4) We encode the sequence f(S) using the Huffman encoding (uniquely decipherable and instantaneous code).

5) We compare the coding limit $NH_0(S)$ of the generated sequence with the length of the encoded transformed sequence f(S).

These steps are repeated a statistically significant number of time and the average value of $NH_0(S)$ and the average values of the length of the encoded transformed sequence f(S) are obtained.

The table shows data obtained with the algorithm described for different values of |A| and N=60.

The first column reports the parameter |A| which indicates the number of symbols of the random sequences generated. The second column reports the average value of the zero-order empirical entropy multiplied by the length of the generated sequence $NH_0(S)$. The third column reports the average value of the number of bits of the transformed sequence f(S) encodes using a code obtained by applying Huffman coding.

All values reported are average values calculated by generating one million sequences.

Contact author: aida.koch445@outlook.com

| \|A\| | Average value $NH_0(S)$ bit | Average value of the length of the encoded transformed sequence f(S) bit |
|---|---|---|
| 30 | 270.4 | 266.9 |
| 35 | 270.3 | 275.4 |
| 40 | 286.0 | 282.2 |
| 45 | 292.8 | 287.7 |
| 50 | 297.5 | 292.4 |

*Table 1: Results obtained for different settings of the parameter |A|, the average values are calculated on one million generated sequences.*

As can be seen from the data reported, the length of the encoded sequence f(S) appears to be smaller on average than the value of $NH_0(S)$.

The program is public, and anyone can use it and test it.

https://it.mathworks.com/matlabcentral/fileexchange/115590-test-sst-huffman-coding

## Analysis of the theorem regarding the $NH_0(S)$ coding limit

In this paragraph we will analyze theorem number 1 of the article "Large Alphabets and Incompressibility" by Travis Gagie. In the introduction of the chapter "Upper bounds" the author makes the following statement:

> We first rephrase the definition of empirical entropy: For $\ell \geq 0$, the $\ell$th-order empirical entropy of a string $S$ is the minimum self-information per character of $S$ emitted by an $\ell$th-order Markov process. The *self-information* of an event with probability $p$ is $\log(1/p)$. An *$\ell$th-order Markov process* is a string of

So, he is stating that when $\ell = 0$ $H_0(S)$ represents is the minimum self-information per character. Consequently, $NH_0(S)$ represents the coding limit of the sequence S using a uniquely decipherable and instantaneous code.

In support of this important statement there is theorem 1 which we report:

Contact author: aida.koch445@outlook.com

**Theorem 1** *For any string $S \in \{1,\ldots,n\}^m$ and $\ell \geq 0$, we have $H_\ell(S) = \frac{1}{m} \min \{ \log(1/\Pr[Q \text{ emits } S]) : Q \text{ is an } \ell\text{th-order Markov process}\}$.*

**PROOF.** Consider the probability an $\ell$th-order Markov process $Q$ emits $S$. Assume, without loss of generality, that $Q$ first emits $s_1 \cdots s_\ell$ with probability 1. For $\alpha \in \{1,\ldots,n\}^\ell$, let $P_\alpha = p_{\alpha,1},\ldots,p_{\alpha,n}$ be the normalized distribution of the characters in $S_\alpha$, so $H(P_\alpha) = H_0(S_\alpha)$; let $Q_\alpha = q_{\alpha,1},\ldots,q_{\alpha,n}$, where $q_{\alpha,a}$ is the probability $Q$ emits $a$ immediately after an occurrence of $\alpha$. Then

$$\log \frac{1}{\Pr[Q \text{ emits } S]}$$
$$= \log \prod_{i=\ell+1}^{m} \frac{1}{q_{s_{i-\ell}\cdots s_{i-1},s_i}}$$
$$= \sum_{i=\ell+1}^{m} \log \frac{1}{q_{s_{i-\ell}\cdots s_{i-1},s_i}}$$
$$= \sum_{|\alpha|=\ell} \sum_{a \in S_\alpha} \#_a(S_\alpha) \log \frac{1}{q_{\alpha,a}}$$
$$= \sum_{|\alpha|=\ell} |S_\alpha| \sum_{a \in S_\alpha} p_{\alpha,a} \left( \log \frac{p_{\alpha,a}}{q_{\alpha,a}} + \log \frac{1}{p_{\alpha,a}} \right)$$
$$= \sum_{|\alpha|=\ell} |S_\alpha|(D(P_\alpha \| Q_\alpha) + H(P_\alpha))$$
$$\geq \sum_{|\alpha|=\ell} |S_\alpha| H(P_\alpha)$$
$$= H_\ell(S) m \, ,$$

with equality throughout if $P_\alpha = Q_\alpha$ for $\alpha \in \{1,\ldots,n\}^\ell$. □

The fundamental point of the proof is the following inequality:

$$= \sum_{|\alpha|=\ell} |S_\alpha|(D(P_\alpha \| Q_\alpha) + H(P_\alpha))$$
$$\geq \sum_{|\alpha|=\ell} |S_\alpha| H(P_\alpha)$$
$$= H_\ell(S) m \, ,$$

Knowing that the relative entropy between P and Q $D(P_\alpha \| Q_\alpha)$ is greater than or equal to zero, it is demonstrated that the zero-order empirical entropy multiplied by *m* is always less than or equal to $\log 1/P[Q\ emit\ S]$. Since $P[Q\ emit\ S]$ is the probability that

Contact author: aida.koch445@outlook.com

the source emits the sequence $S$, the function $\log 1/P[Q\ emit\ S]$ represents the information content of the message from the point of view of the source.

<u>Therefore, from this inequality it follows that the empirical entropy $H_0(S)$ underestimates or equals the entropy of the source. From here, the theorem reaches the conclusion that $H_0(S)$ represents the minimum information per character.</u>

This result implies that there is no transform that manages on average to reduce the value of $NH_0(S)$. This hypothesis is not supported by any demonstration, and we believe that this point represents a criticality to the reported theorem.

Since these are sequences made up of random variables, there is a common belief that it is not necessary to demonstrate the non-existence of a function capable of reducing the average value of $NH_0(S)$ of the sequence. Because a function of this type is considered as a function capable of making a random sequence a little less random, an obviously impossible operation.

This way of reasoning has two important weaknesses:

1) every single hypothesis must always be proven even if it may seem trivial or stupid.
2) we must be very careful about taking for granted hypotheses that deal with topics that include the concepts of information and randomness, subjects that are not yet fully understood.

<u>In conclusion, the theorem only demonstrates that empirical entropy tends to underestimate the entropy of the source that generated the message. This result is not sufficient to consider empirical entropy as the minimum information per character.</u>

By applying the set shaping theory, as demonstrated in our counterexample, it is possible to reduce the value of $NH_0(S)$ on average. Consequently, the hypothesis underlying this theorem turns out to be incorrect.

Based on this consideration, the theorem it just proves that $H_0(S)$ represents the minimum self-information per character under the condition that the frequencies of the symbols in S are kept unchanged, therefore not applying any transform.

In order to define minimum self-information per character, we must consider all the possible transforms of S, this important step is not present in the reported theorem.

Our counterexample uses a source that will generate a random sequence of symbols emitted uniformly. Therefore, this source represents a zero-order Markovian process that falls within the conditions defined by the theorem.

Contact author: aida.koch445@outlook.com

**Theorem 1** *For any string $S \in \{1,\ldots,n\}^m$ and $\ell \geq 0$, we have $H_\ell(S) = \frac{1}{m} \min \{ \log(1/\Pr[Q \text{ emits } S]) : Q \text{ is an } \ell\text{th-order Markov process} \}$.*

Therefore, according to the theorem, it is impossible to encode on average the sequence S, generated by a Markovian process of order 0 l=0, using a uniquely decipherable code with less than $NH_0(S)$ bits. Our counterexample demonstrating that this is possible. Therefore, demonstrates that the zero-order empirical entropy does not represent the minimum information of the character. Consequentially, NH0(S) does not represent the average coding limit of the symbols of an individual sequence S of random variables i.i.d. through a uniquely decipherable and instantaneous code.

## Conclusions

The purpose of this article is not to give answers but to pose a problem to the scientific community regarding the meaning given to the parameter $NH_0(S)$ which is considered the coding limit of an individual sequence. The counterexample presented is incompatible with this limit. We also analyzed a theorem that proves that $H_0(S)$ represents the minimum self-information per character, showing that this theorem is based on a fundamental hypothesis that is not proven. This hypothesis is based on the common opinion that there are no transforms capable of reducing the value of $NH_0(S)$ on average, unfortunately there are no demonstrations to support this hypothesis. The set shaping theory shows that this hypothesis is wrong, there is a transform that manages to reduce the value of $NH_0(S)$ on average, without changing the alphabet.

Given the enormous importance that this limit has in information theory, we believe that the time has come for the scientific community to address this problem.

## Reference

1. C Schmidt, A Vdberg, A Petit, "Practical applications of Set Shaping Theory in Huffman coding", arXiv preprint arXiv:2208.13020, 2022.
2. T Gagie , "Large alphabets and incompressibility" Information Processing Letters, 2006 – Elsevier.
3. Cover, Thomas M. (2006). Elements of Information Theory. John Wiley & Sons. ISBN 0-471-24195-4.
4. Shannon, C. E., & Weaver, W. (1949). The mathematical theory of communication. Urbana: University of Illinois Press.
5. Solomon Kozlov. Introduction to Set Shaping Theory. ArXiv, abs/2111.08369, 2021.
6. Solomon Kozlov. Use of Set Shaping theory in the development of locally testable codes. arXiv:2202.13152. February 2022.

Contact author: aida.koch445@outlook.com

7. Biereagu, Sochima. 2023. "Introducing the Role of Shaping Order K in Set Shaping Theory." AfricArXiv. October 12. doi:10.31730/osf.io/ywmsr.
8. Huffman, D. (1952). A Method for the Construction of Minimum-Redundancy Codes. Proceedings of the I.R.E. 40 (9):1098 1101. doi: 10.1109/JRPROC.1952.273898.

Contact author: aida.koch445@outlook.com